\documentclass[12pt]{amsart}         
\usepackage{graphicx}        
\usepackage{amsmath,amsthm,amssymb,bm}
\usepackage{commath,mathtools,mathrsfs}
\usepackage[font=tiny,labelfont=bf]{caption}
\usepackage{hyperref}
\hypersetup{
    colorlinks,
    citecolor=blue,
    filecolor=black,
    linkcolor=black,
    urlcolor=black
}
\usepackage{bbm,color}
\usepackage{hyperref}

\calclayout

\def\R{\mathbb R}

\def\E{\mathbb E}

\newtheorem{thm}{Theorem}[section]

\newtheorem{defi}[thm]{Definition}

\graphicspath{{images/}} 

\begin{document}
\title{Portfolio Diversification Revisited}

\thanks{email: charles@fixedpoint.io}
\author{Charles Shaw \\ FixedPoint IO Ltd.}
\date{This version: 27 April 2022.}

 \maketitle

\begin{abstract}
We relax a number of assumptions in Alexeev and Tapon (2012) in order to account for non-normally distributed, skewed, multi-regime, and leptokurtic asset return distributions. We calibrate a Markov-modulated L\'evy process model to equity market data to demonstrate the merits of our approach, and show that the calibrated models do a good job of matching the empirical moments. Finally, we argue that much of the related literature on portfolio diversification relies on assumptions that are in tension with certain observable regularities and which, if ignored, may lead to underestimation of risk. 
\end{abstract}

\vskip5mm


\newpage
\tableofcontents
\newpage

\section{Introduction}

It is by now generally accepted that, absent inside information, diversification is smart and that unsystematic (idiosyncratic) risk, or the risk that is peculiar to one firm or industry, may be reduced by diversification. Evidence suggests that idiosyncratic risk is the most significant contributor to overall volatility, and these two components are not totally independent. Due to this, it is impossible to totally eliminate risk even with a high number of investments. Even with international diversification, there will always be a portion of total risk that is related to systemic risk factors. 

This and other related questions are remain an active topic of research discussion, both among investment industry practitioners and academic researchers. Simply stated, the challenge is to invest an initial sum of money in financial assets in order to maximize the expected utility of the terminal wealth. In an early attempt to address this question, Markowitz \cite{M52} devised the mean-variance approach to portfolio optimization and pioneered the application of quantitative approaches for the question optimum portfolio selection. Black, Scholes, and Merton \cite{Merton76, B73} promoted an extension of these results and provided explicit formulation of portfolio asset selection and allocation in continuous time. Despite the fact that Merton et al's techniques deliver substantial theoretical conclusions, they yield some practical concerns. One such concern stems from the assumption that a risky asset's dynamics follow a geometric Brownian motion. Some empirically observable regularities of financial time-series, such as the asymmetry and heavy-tailedness of the distribution of returns of time-varying conditional volatility, are in tension with this approach.

Another limitation of the Black-Scholes-Merton framework is that the coefficients are static. This appears to be an important consideration when it comes to longer-term investment horizons, when macroeconomic conditions may plausibly shift several times and fundamentally alter investment opportunity sets. In this context, Markov-modulated -- otherwise called \textit{regime switching} -- models appear to be a good fit for representing these kinds of phenomena. For example, changes in the condition of the economy may be reasonably\footnote{We will return to expand on what "reasonable" could mean in this context when we introduce the Regime Classification Measure later in this paper.} quantified using a Markov chain, which modify the parameters of the model. These models may therefore be used to describe macroeconomic shifts, times when market behaviour changes dramatically during crisis episodes, or the various stages of business cycles.  

When considering portfolio optimisation, there are a slew of variables to consider when trying to figure out how many assets are necessary to achieve optimum diversification. For instance, systematic risk could be measured in a variety of ways, and could depend on the size of the investment universe, the investor's characteristics, asset features (which may be time-varying), the model used to measure diversification, the frequency of the data that is used, market conditions, the time horizon, and other such parameters. One could argue that calculating an optimal number of assets a fully-diversified portfolio -- for a given market, time horizon, or a set of preferences -- is fraught with problems if the issue of asset allocation is considered from a classic Mean-Variance Optimisation perspective \`a la Markovitz. Nevertheless, recent research have attempted to argue that the size of a well-diversified portfolio is greater now than it has ever been, that this number is smaller in developing economies compared to established financial markets, and this number decreases with increasing stock correlations with the market \cite{ATa12, AD15, C+01}. 

In this paper, we seek to critically explore these and closely related themes in the following manner. By deploying a Markov-modulated L\'evy process, we first relax a number of assumptions that are essentially "baked in" in studies such as Alexeev and Tapon \cite{ATa12} in order to account for account for non-normally distributed, skewed, multi-regime, and leptokurtic asset return distributions. We calibrate a model to equity market data and show that the calibrated models do a good job of matching the empirical moments. We then argue that much of the related literature on portfolio diversification relies on assumptions that are in tension with certain observable regularities and which, if ignored, may lead to material underestimation of risk.

The rest of this paper is organised as follows. Section \ref{sec:lit} provides a review of the related literature, section \ref{sec:data} provides a calibration exercise where a Markov-modulated L\'evy process is fitted to Nasdaq data. Here, we relax a number of assumptions in Alexeev and Tapon \cite{ATa12} in order to account for possibly non-normally distributed, skewed, multi-regime, and leptokurtic asset return distributions. Section \ref{sec:discussion} provides a discussion of the results and section \ref{sec:conclusion} concludes.

\section{Literature review}
\label{sec:lit}
Classic portfolio allocation theory is based on the mean–variance (MV) framework, which gives a framework to analyse the trade-off between risk and return for attaining diversity advantages. Despite this framework's limitations, asset allocation choices are frequently based on it as many asset managers, consultants, and investment advisers employ classical MV optimisation as a standard quantitative method for portfolio construction. This framework stems from the work of Markowitz \cite{M52} who in 1952 developed the Modern Portfolio Theory, laying the groundwork for the subsequent development of risk and return theories. Along with Markowitz, Evans and Archer \cite{EA68}, Fielitz \cite{F74}, Solnik \cite{S74}, Statman \cite{S87}, and Campbell et al. \cite{C+01} are some of the pioneering writers in the field of portfolio risk diversification. Their results can be summarised as follows:

\begin{itemize}
\item A portfolio's risk may be decreased by diversification, which includes both systematic and non-systematic risk. 
\item A portfolio's overall risk increases when the number of stocks it holds approaches that of a market's total. 
\item Unsystematic risk may be minimised up to the point of achieving optimal portfolio diversification - meaning that the total portfolio risks are equivalent to their systematic counterparts.
\end{itemize}

Before elaborating further, let us first examine some of the key issue around measurement of risk diversification. Portfolio risk may be measured in a variety of ways, each with their own set of merits and demerits, and which have been the subject of continued debate. Standard deviation tends to be used in many investigations as a commonly acknowledged metric of risk. This is not only evident in early studies (\cite{EA68,S74,S87,BPP96}), which all addressed standard deviation in their research. This pattern is also present in more recent research (\cite{BG05}, \cite{B10}). One of the main problems of using standard deviation as a risk metric is that it might lead to erroneous and misleading results since the metric is particularly sensitive to extreme values and outliers. It is well-known that the standard deviation may lead to inaccurate estimates of extreme occurrences if the returns are not Gaussian. Another problem with using a naive standard deviation approach is that both positive and negative variations from the average return are treated equally. With this in mind, a considerable amount of research has been done on the potential of expected shortfall (ES) and terminal wealth standard deviation (TWSD) as an extreme risk measure to examine the impact of the financial crisis on the optimal number of stocks in the portfolio (\cite{BG05, B10}). As an alternative measure of portfolio risk, mean absolute deviation (MAD) \cite{F74} employs absolute deviation rather than variance, while the unsystematic risk ratio (URR) \cite{SV18} provides a measure of diversification relative to its variation. From one risk measure to the next, the portfolio structures generated are vastly different.

The heterogeneity of viewpoints on what constitutes successful diversification makes it difficult to reliably compare work across the literature on optimum portfolio diversification. For example, Alexeev and Dungey \cite{AD15} claim that investing in a diversified portfolio of seven (10), evenly weighted equities may reduce risk by as much as 85 per cent (90 per cent). But a portfolio of just 20 equities is needed to remove 95 per cent of the unsystematic risk, while a portfolio of 80 equities may reduce an extra 4 per cent of unsystematic risk, according to Tang \cite{T04}. Alekneviciene et al \cite{AAR12} argue that a differently-weighted portfolio could remove 97 per cent of unsystematic risk with 25 stocks, while Stotz and Lu \cite{SL14} found that in China, 67 per cent of unsystematic risk could be managed away with just 10 stocks. According to Kryzanowski and Singh \cite{KS10}, investing in 20–25 stocks may minimise 90 per cent of a portfolio's risk. They find that investors who are risk-averse prefer a portfolio with a risk reduction of 99 per cent, while more aggressive investors seeking bigger returns at the expense of more risk may be satisfied with a portfolio risk reduction of 90 per cent. Alexeev and Tapon \cite{ATa12} further argue that the required number of stocks in a well-diversified portfolio is determined by the average correlations between stocks and the market circumstances, whether they are distressed or quiescent, when analysing the dynamics of portfolio holdings over multiple years. Raju and Agarwalla \cite{RA21} claim that the appropriate number of stocks is influenced by the investor's risk tolerance and desired degree of confidence, as well as the portfolio's weighting structure. These findings make intuitive sense, since investors' actions vary according to their economic circumstances. Most of the time, their actions are dynamic, heavily impacted by economic, cultural, and social variables, and not necessarily reasonable \cite{K16}. As a result, their values, tastes, assumptions, and preconceptions shift in tandem with the economy \cite{AB17}. Additionally, the investor's location has an impact on both their optimum asset allocation and their investment success \cite{FW06}.

It is also possible that the frequency of the data utilised might have an effect on the ideal number of stocks. This may lead to an exaggerated number of stocks in a portfolio, as shown by Alexeev and Dungey \cite{AD15}. They also point out that this disparity is magnified during financial market crises. The optimal number of equities for optimal diversification is 8 to 16, according to early research based on (semi)annual and quarterly data \cite{EA68, F74, Z14}. High-frequency data, as Alexeev and Dungey \cite{AD15} further argue, significantly enhances risk assessment and decision-making. Using higher frequency data, the number of stocks needed to accomplish the desired risk reduction drops. Data from various frequency intervals showed a minor variation in unsystematic risk during calm times, but a significant difference during periods of high volatility. The estimates of diversifiable risk were shown to be overstated when lower frequency data were employed, particularly during times of financial difficulty. Risk measurements based on more frequent data beat those based on less frequent data, therefore holding big portfolios is not always required, as indicated by the lower frequency risk measures, particularly during a financial crisis. Financial instruments' price fluctuations are also impacted by basic variables like interest rates, economic development, currencies, and so forth. The diversification advantages of constructing a portfolio based on these criteria might be discovered by investors \cite{K16}. Corporate bonds, in addition to stocks, may provide considerable risk reduction, particularly in times of financial turbulence.

Following this line of thinking, an aggregate stock index built of the S\&P 500's most prominent companies may cut transaction costs while providing enough equities for diversification, according to Aboura and Chevallier \cite{AC17}. To prevent international financial contagion, investors and mutual fund managers alike need to diversify their portfolios and use hedging methods \cite{F+18}; funds with less liquid equities in their portfolios tend to be more diversified, according to these findings. According to Pastor et al. \cite{P+20}, they were found to be larger and cheaper, as well as trading more often. Active portfolio management has taken another step forward with this development. Socially responsible investment funds are determined to have no influence on idiosyncratic risk due to screening intensity, which is an essential aspect of an investing strategy \cite{L+10}. Moreover, managers may prefer to shift their portfolios during moments of political instability in favour of firms that report more accurately \cite{C+18,CVP18}.

There are several difficulties in trying to model financial time series since the observations might be affected by events that are generally unforeseen. Events such as natural catastrophes, statements from central banks, and policy announcements from governments may have a significant impact on the market. It is therefore quite possible that the assumption of stationary financial data is violated. As a result of this violation, classic approaches to time series analysis may not be satisfactory. Markov-switching models are therefore of interest since, under certain mild assumptions, they allow us to mitigate issues around suspected non-stationarity of time series.

An important stylized feature of financial time series is regime switching i.e. when economic conditions undergo periodic regime changes. The regular flow of economic activity may occasionally suffer shocks substantive enough to result in different observed dynamics. This means that sampled time series data may typically show not only periods of low and high volatility but also periods of slower and faster mean growth. The economy may oscillate between two states: 1) a stable, low-volatility state characterised by economic expansion, and 2) a panic-driven, high-volatility state defined by economic contraction. Such dynamics are specified by deterministic or stochastic ordinary differential equations. Regime shifts indicate transitions from one kind of dynamics to another, with the possible change in state space, the objective function, or both. Regime changes may occur for a variety of reasons: (i) exogenous changes in dynamics (for example, as a result of sudden environmental disasters or social/political reform); (ii) unintentional internal changes in dynamics (for example, as a result of human activity-related disasters or firm bankruptcy); (iii) intentional (controlled) shifts to new dynamics (technological innovations, mergers of firms, etc.); (iv) changes in preferences/objectives (environmental concerns. A single model may include a combination of the above-mentioned triggers. Evidence of such regimes has been widely documented and switching models have since been used in the literature across various domains of application, including but not limited to exchange rates, asset allocation, and equity markets. Surveys are provided by \cite{AT12} and \cite{H16}.

\section{Data and model}
\label{sec:data}

\begin{figure}[htbp!]
\centering
\caption{Log returns for Nasdaq 100 (blue) with with normal distribution overlay (red)} 
\label{fig:sp500.1}
\centering
\includegraphics[width=1\textwidth]{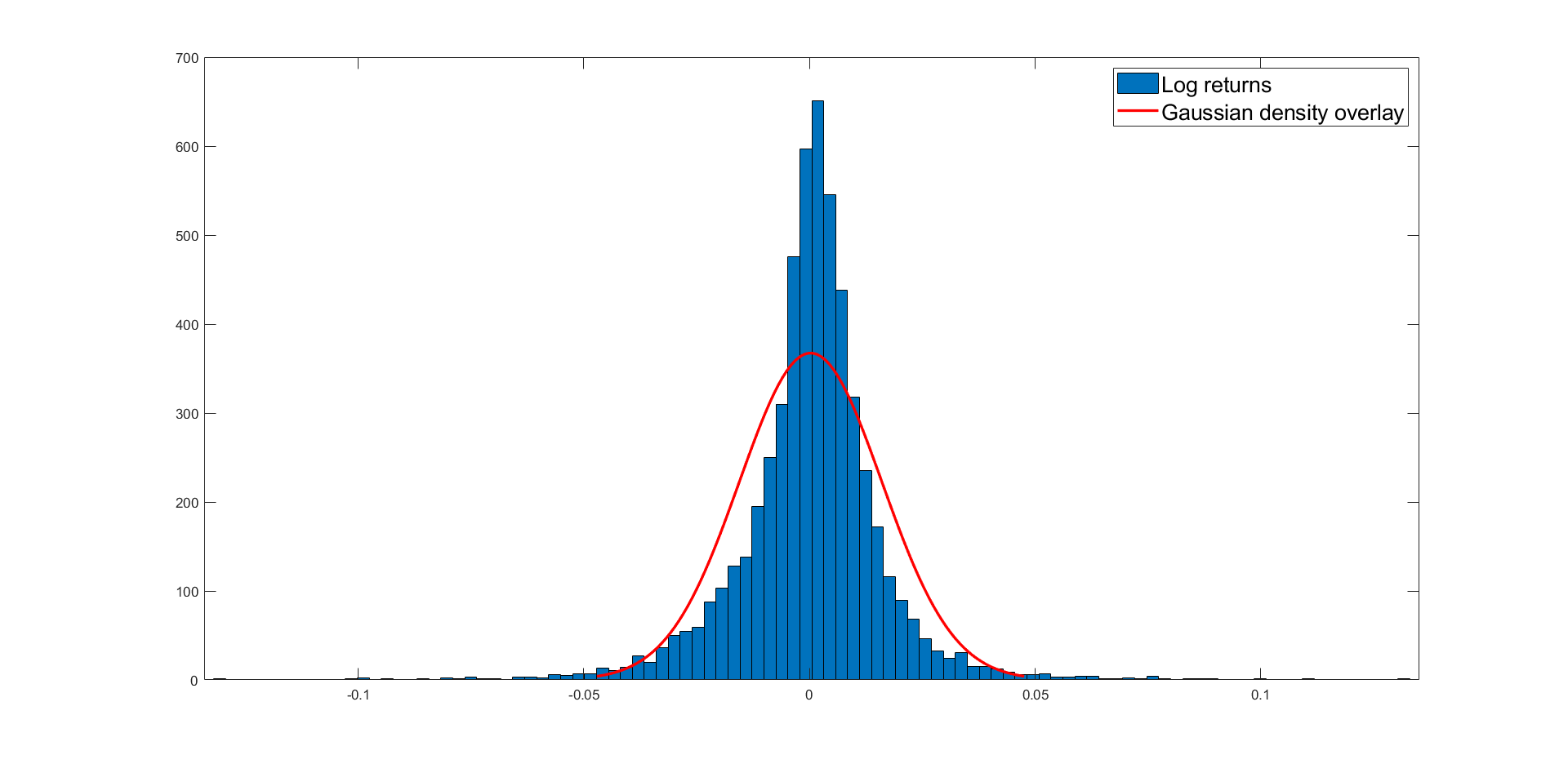}\hfill
\end{figure}

To facilitate the discussion, we will use use daily closing price data for the Nasdaq 100 index, taken from Oxford-Man's Realized Library\footnote{https://realized.oxford-man.ox.ac.uk}. Our data spans from 04 Jan 2000 - 12 Nov 2021, resulting in 5483 daily observations. Figure \ref{fig:sp500.1} shows the plot of log returns for Nasdaq 100 (blue) with with normal distribution overlay (red). The graph clearly shows that Gaussian density underestimates random variation. This can be explained by the fact that financial returns tend to have distributions with certain stylized features, such as excess kurtosis (i.e. fat or semi-heavy tails), skewness (tail asymmetries), and regime shifts.  

Practical modelling of said stylized features may be achieved by the General Hyperbolic Distribution. The special case of the General Hyperbolic Distribution on which we focus our attention is the Normal-Inverse Gaussian (NIG) distribution, a continuous probability distribution defined as the normal variance-mean mixture with an inverse Gaussian mixing density \cite{B97,BN04}. The four-dimensional parameter vector $[\alpha, \beta, \mu, \delta]$ specifies the form of the NIG density. Due to the extensive parametrization, the NIG density is an advantageous model for a wide range of unimodal positive kurtotic data. The alpha-parameter determines the density's steepness or pointiness, which rises monotonically as $\alpha$ increases. A high alpha value indicates that the tails are light, whereas a small number indicates that the tails are heavy. The beta-parameter is used to specify the skewness. When $\beta<0$ ($\beta>0$), the density is skewed to the left, when $\beta>0$, it is skewed to the right, and when $\beta = 0$, the density is symmetric around $\mu$, which is a centrality parameter. The delta-parameter is a scale parameter. The class of NIG distributions is a flexible system of distributions that includes fat-tailed and skewed distributions, and the normal distribution, $N(\mu ,\sigma ^{2})$, arises as a special case by setting  $\beta =0,\delta =\sigma ^{2}\alpha $, and letting $\alpha \rightarrow \infty$ .

Consider the following generic Markov Regime Switching (MRS) model:
\begin{equation}\label{MSM}
\begin{cases}
y_t = f(S_t,\theta,\psi_{t-1})\\
S_t = g\left(\tilde{S}_{t-1},\psi_{t-1}\right) \\
S_t \in \Lambda
\end{cases}
\end{equation}
where $\theta$ is the vector of the parameters of the model, $S_{t}$ is the state of the model at time $t$, $\psi_t:=\left\{y_k: k=1,\dots,t\right\}$ is the set of all observations up to $t$, $\tilde{S}_{t} := \{S_{1},...,S_{t}\}$ is the set of all observed states up to $t$, $\Lambda=\{1,...,M\}$ is the set of all possible states, and $g$ is the function that regulates transitions between states. Function $f$ indicates how observations at time $t$ depend on $S_t,\theta,\ \text{and}\ \psi_{t-1}$ and finally, $t \in \{0,1,...,T\}$, where $T \in \mathbb{N}$, $T < +\infty$, is the terminal time.

Equations \ref{MSM} allow us to address specific issues that may be difficult to represent in a single state regime, which is clearly useful for time series applications. Although the literature on Markov-switching models is diverse, two general groups may be identified. The first group consists of models that have basic transition laws (such as a first order Markov chain) but complex distributions for the data or a high number of states. For examples of research in this area see \cite{DiPMF16,Ham89,Ch98}. The second group consists  of models with more complicated transition laws but with simple assumptions and few states, often limited to two. See, for example, \cite{Kim99, DiLeeW94, Pe05}.

\label{sec:stoch}
We now turn our attention to estimation of a Markov-switching model augmented by jumps, under the form of a L\'evy process, with a view to applying this methodology to model financial returns. In order to motivate this section's modelling approach, we first set up the general structure of L\'evy processes, then we outline their properties with reference to path variation and the L\'evy-Khintchine theorem. The discussion highlights the properties of Markov chain, such as irreducibility, aperiodicity, and ergodicity. The section includes a discussion on a framework for estimating a jump-robust model tempered by a Markov chain, which can be used to study the relations of dependence within financial returns. 

\subsection{L\'evy processes}
L\'evy processes can be thought of as a combination of two distinct processes, namely diffusions and jumps. The attractive properties of such a combination can demonstrated by sketching the connections between two. A well-known pure diffusion process used in finance is the Wiener process, a continuous-time Markovian stochastic process with a.s. continuous sample paths. A well-known pure jump process is the Poisson process, which is a non-decreasing process that, unlike Wiener, does not have continuous paths. Whilst the Poisson process has paths of bounded variation over finite time horizons, the paths of a Wiener process exhibit unbounded variation over finite time horizons. 

When combined, these become interesting and, crucially, tractable tools for modelling financial time series due to their ability to match the empirically observed behaviour of financial markets more accurately than when armed with simple Wiener process-based models. These tools are useful, for example, in modelling jumps, spikes, and other such discontinuous variations in the price signal that are frequently observed in asset prices processes. Such jump dynamics may be due to short-term liquidity challenges, microstructure frictions, or news shocks. Despite their apparent differences, these two processes have much in common. Both processes are initiated from the origin, both have right-continuous paths with left limits\footnote{ We adopt the convention that all L\'evy processes have sample paths that are cadlag or RCLL i.e. right-continuous with left limits at every $t$.}, and both have independent and stationary increments. Hence, these common features can be generalised to define a common framework of one-dimensional stochastic (L\'evy) processes. 

\subsection{A regime-switching L\'evy model}
This subsection motivates the introduction of the regime-switching L\'evy approach to modelling of our time series. This part of the discussion follows the methodology proposed by \cite{CG16}, namely by combining a L\'evy jump-diffusion model with a Markov-switching framework. The regime-switching L\'evy model offers the possibility of identification of such stochastic jumps, together with disentangling different market regimes and capturing the regime-switching dynamics. We begin by introducing a number of key definitions and notations. 

\begin{defi}[Stochastic Process]
A stochastic process $X$ on a probability space $(\Omega, \mathcal{F}, \mathbb P)$ is a collection of random variables $(X_t)_{0 \leq t < \infty}$.
\end{defi}
If $X_t \in \mathcal F_t$, the process $X$ is adapted to the filtration $\mathcal{F}$, or equivalently, $\mathcal F_t$-measurable.
\newline
\begin{defi}[Brownian Motion] \label{3}
Standard Brownian motion $W=(W_t)_{0 \leq t < \infty}$ has the following three properties:\\
\noindent (i) $W_0=0$\\
(ii) $W$ has independent increments: $W_t-W_s$ is independent of $\mathcal{F}_s,$  $0 \leq s<t< \infty$\\
(iii) $W_t-W_s$ is a Gaussian random variable: $W_t-W_s \sim N(0,t-s)$ $\forall$ $0 \leq s < t < \infty$
\end{defi}
Property $(ii)$ implies the Markov property i.e. conditional probability distribution of future states of the process depend only on the present state. Property $(iii)$ indicates that knowing the distribution of $W_t$ for $t \leq \tau$ provides no predictive information about the state of the process when $t > \tau$. We can also define a Poisson Process, another stochastic mechanism as follows. 
\newline
\begin{defi}[Poisson Process] \label{poiss}
A Poisson process $N=(N_t)_{0 \leq t < \infty}$ satisfies the following three properties:\\
\noindent (i) $N_0=0$\\
(ii) $N$ has independent increments: $N_t-N_s$ is independent of $\mathcal{F}_s,$  $0 \leq s<t< \infty$\\
(iii) $N$ has stationary increments:  $P(N_t-N_s \leq x)=P(N_{t-s} \leq x),$    $ 0 \leq s<t< \infty $ 
\end{defi}

SDEs formulated with only the Poisson process or Brownian motion may not be very useful in investing or risk management. Arguably one needs more realistic models to describe the complex dynamics of an evolving system. However, their common properties may be combined, thus establishing a more general process. 
\newline
\begin{defi}[L\'evy Process] \label{Levy}
Let $L$ be a stochastic process. Then $L_t$ is a L\'evy process if the following conditions are satisfied:\\
\noindent (i) $ L_0=0$\\
(ii) $L$ has independent increments: $L_t-L_s$ is independent of $\mathcal{F}_s,$  $0 \leq s<t< \infty$\\
(iii) $L$ has stationary increments:  $\mathbb{P}(L_t-L_s \leq x)=\mathbb{P}(L_{t-s} \leq x),$    $ 0 \leq s<t< \infty $ \\
(iii) $L_t$ is continuous in probability: $\lim_{t \to s} L_t = L_s$
\end{defi}

Condition (iii) follows from (i) and (ii). For proof see \cite{K97}.
\newline
\begin{defi}
A real valued random variable $\Theta$ has an infinitely divisible distribution if for each $n=1,2, \ldots$, there exists a $i.i.d.$ sequence of random variables $\Theta_1,\ldots, \Theta_n$ such that
\[\Theta \,{\buildrel d \over =}\,\Theta_{1,n}+\ldots+\Theta_{n,n} \]
This says that the law $\mu$ of a real valued random variable is infinitely divisible if for each $n = 1, 2, \ldots$ there exists another law $\mu_n$ of a real valued random variable such that $\mu=\mu_n^{*n}$, the n-fold convolution of $\mu_n$.
\end{defi}

The full extent to which we may characterize infinitely divisible distributions is carried out via their characteristic function (or Fourier transform of their law) and the L\'evy-Khintchine formula. 

\begin{thm}[L\'evy-Khintchine formula]\label{LK}
Suppose that $\mu \in \R$, $\sigma \geq 0$, and $\Pi$ is a measure concentrated on $\R / \{0\}$ such that $\int_\R \min(1,x^2) \Pi(dx) < \infty$. A probability law $\mu$ of a real-valued random variable $L$ has characteristic exponent $\Psi(u):= - \frac{1}{t} \log \E [e^{iuL_t} ]$ given by,
\begin{align}
\Phi(u;t)=\displaystyle\int_\R e^{i u x} \mu (dx) = e^{- t\Psi(u)} \quad \text{for}\quad u \in \R,
\end{align}
if (and only if) there exists a triple $(\gamma, \sigma,\Pi)$, where $\gamma \in \mathbb{R}, \sigma \geq 0$ and $\Pi$ is a measure supported on $\mathbb{R}\setminus\{0 \}$ satisfying  $\int_{\mathbb{R}} (1\wedge x^2) \Pi (dx) < \infty$, such that 
\begin{align}
\Psi(\lambda) = i \gamma u + \frac{\sigma^2 u^2}{2} + \int_{\mathbb{R}} \left( 1-e^{(iux)} +iux \mathbf{1}_{|x|<1} \right) \Pi(dx) 
\end{align}
for all $u \in \R$.
\end{thm} 

From Theorem \ref{LK} we can say that there exists a probability space where $L=L^{(1)}+L^{(2)}+L^{(3)}$. We can build intuition as follows: $L^{(1)}$ is standard Brownian motion with drift, $L^{(2)}$ is a compound Poisson process, and $L^{(3)}$ is a square integrable martingale with countable number of jumps of magnitude less than unity (a.s.). This is the the L\'evy-It\^{o} decomposition, which can be stated as follows
\begin{align} \label{itodecomp}
L_t = \eta t + \sigma W_t + \int\limits_0^t \int\limits_{|x| \geq 1} x \mu^L(ds,dx) + \int\limits_0^t \int\limits_{|x|<1} x(\eta^L - \Pi^L)(ds,dx).
\end{align}

\begin{defi}[Markov-Switching] \label{MS}
Let $(Z_t)_{t \in [0,T]}$ be a continuous time Markov chain on finite space $S:=\{1, \ldots,K\}$. Let $\mathcal{F}^Z_t:=\{\sigma(Z_s); 0 \leq s \leq t \}$ be the natural filtration generated by the continuous time Markov chain $Z$. The generator matrix of $Z$, denoted by $\Pi^Z$, is given by 
\begin{equation}
    \Pi^Z_{ij}
\begin{dcases}
   \quad  \geq 0,                        & \text{if } i \neq j\\
     - \sum_{j\neq i} \Pi^Z_{ij},   & \text{otherwise}
\end{dcases}
\label{eq2}
\end{equation}
\end{defi}
We can now define the Regime-switching L\'evy model as follows.
\newline
\begin{defi}[Regime-switching L\'evy model] For all $t \in [0,T]$, let $Z_t$ be a continuous time Markov chain on finite space $S:=\{1,\ldots, K\}$ defined as per Definition \ref{MS}. A regime-switching model is a stochastic process $(X_t)$ which is solution of the stochastic differential equation given by 
\begin{equation}
\label{SDE}
dX_t=k(Z_t)(\theta(Z_t)-X_t)dt+\sigma(Z_t)dY_t 
\end{equation}
where $k(Z_t)$, $\theta(Z_t)$, and $\sigma(Z_t)$ are functions of the Markov chain $Z$. They are scalars which take values in $k(Z_t)$, $\theta(Z_t)$, and:
\begin{align*}
& \sigma(Z_t): k(Z_t):=\{k(1),\ldots,k(K)\} \in \mathbb{R}^{K^*}, \\ 
& \theta(S):=\{\theta(1),\ldots,\theta(K)\}, \\
& \sigma(S):=\{\sigma(1),\ldots,\sigma(K)\} \in \mathbb{R}^{K^*},
\end{align*}
where $Y$ is a Wiener or a L\'evy process. Here, $k$ denotes the mean reverting rate, $\theta$ denotes the long run mean, and $\sigma$ denotes the volatility of $X$.
\end{defi}

The above model exhibits two sources of randomness: the Markov chain $Z$, and the stochastic process $Y$ which appears in the dynamics of $X$. In other words, there is stochasticity due to the Markov chain Z, $\mathcal{F}^Z$, and stochasticity due to the market information which is the initial continuous filtration $\mathcal{F}$ generated by the stochastic process Y.

\subsection{NIG-type distribution}
Following \cite{CG16}, let us assume that the L\'evy process $L$ follows the Normal Inverse Gaussian (NIG) distribution, defined as a variance-mean mixture of a normal distribution with the inverse Gaussian as the mixing distribution (also see Barndorff-Nielsen et al \cite{B97, BN01, BNX1, BNX2} for a richer discussion on this topic). 

The NIG type distribution is a relatively novel process introduced by Barndorff-Neilsen \cite{BN01} as a model for log returns of stock prices. It is a sub-class of the more general class of hyperbolic L\'evy processes. After its introduction it was demonstrated that the NIG distribution provides an excellent fit to log returns of stock market data \cite{BB01}. Other studies have also shown this distribution's superior empirical fit to other asset classes \cite{Prause3, Kal, NIG1, NIG2}. \cite{Suar} find that the Normal Inverse Gaussian distribution provides an overall fit for the data better than any of the other subclasses of Generalized Hyperbolic distributions and much better than the L\'evy-stable laws. Rachev et al \cite{Rach} have deployed the NIG distribution to successfully address such well-known 'puzzles' as (i) predictability of asset returns (ii) the equity premium, and (iii) the volatility puzzle. 

This type of types of heavy-tailed process is of interest, particularly since the NIG distribution fulfils the fat-tails condition, is analytically tractable, yet is closed under convolution \cite{ NIG3}. 

The density function of a $NIG(\alpha, \beta, \delta, \mu)$ is given by 
\begin{equation}
f_{NIG}(x;\alpha, \beta, \delta, \mu)= \frac{\alpha}{\pi} e^{\delta~ \sqrt[]{\alpha^2-\beta^2+\beta(x-\mu)}} \frac{K_1(\alpha \delta~ \sqrt[]{1+(x-\mu)^2/\delta^2})}{\sqrt[]{1+(x-\mu)^2/\delta^2}},
\end{equation}
where $\delta>0$, $\alpha \geq 0$. The parameters in the Normal Inverse Gaussian distribution can be interpreted as follows: $\alpha$ is the tail heaviness of steepness, $\beta$ is the skewness, $\delta$ is the scale, $\mu$ is the location. The NIG distribution is the only member of the family of general hyperbolic distributions to be closed under convolution. $K_v$ is the Hankel function with index $v$. This can be represented by 
\begin{equation}
K_v(z) =\frac{1}{2} \int^\infty_0 y^{v-1} e^{\Big(-\frac{1}{2}z \left(y+\frac{1}{y} \right)  \Big)} dy
\end{equation}
For a given real $v$, the function $K_v$
satisfies the differential equation given by
\begin{equation}
v^2 y''+xy'-(x^2+v^2)y=0.
\end{equation}
The log cumulative function of a Normal Inverse Gaussian variable is given by 
\begin{equation}
\phi^{NIG}(z)=\mu z +\delta ~ (\sqrt[]{\alpha^2-\beta^2}-\sqrt[]{\alpha^2-(\beta+z)^2)} \textnormal{~~for~all~} |\beta+z|<\alpha.
\end{equation}

The first two moments are $\mathbb{E}[X]=\mu+\frac{\delta \beta}{\gamma}$, and $Var[X]=\frac{\delta \alpha^2}{\gamma^3}$, where $\gamma=\sqrt[]{\alpha^2-\beta^2}$. The L\'evy measure of  a $NIG (\alpha, \beta, \delta, \mu)$ law is 
\begin{equation}
F_{NIG}(dx)=e^{\beta x} \frac{\delta \alpha}{\pi |x|} K_1 (\alpha|x|) dx.
\end{equation}

\begin{figure}[htbp!]
    \fbox{%
    \begin{minipage}[t]{0.47\linewidth}
        \centering
\includegraphics[width=1\textwidth]{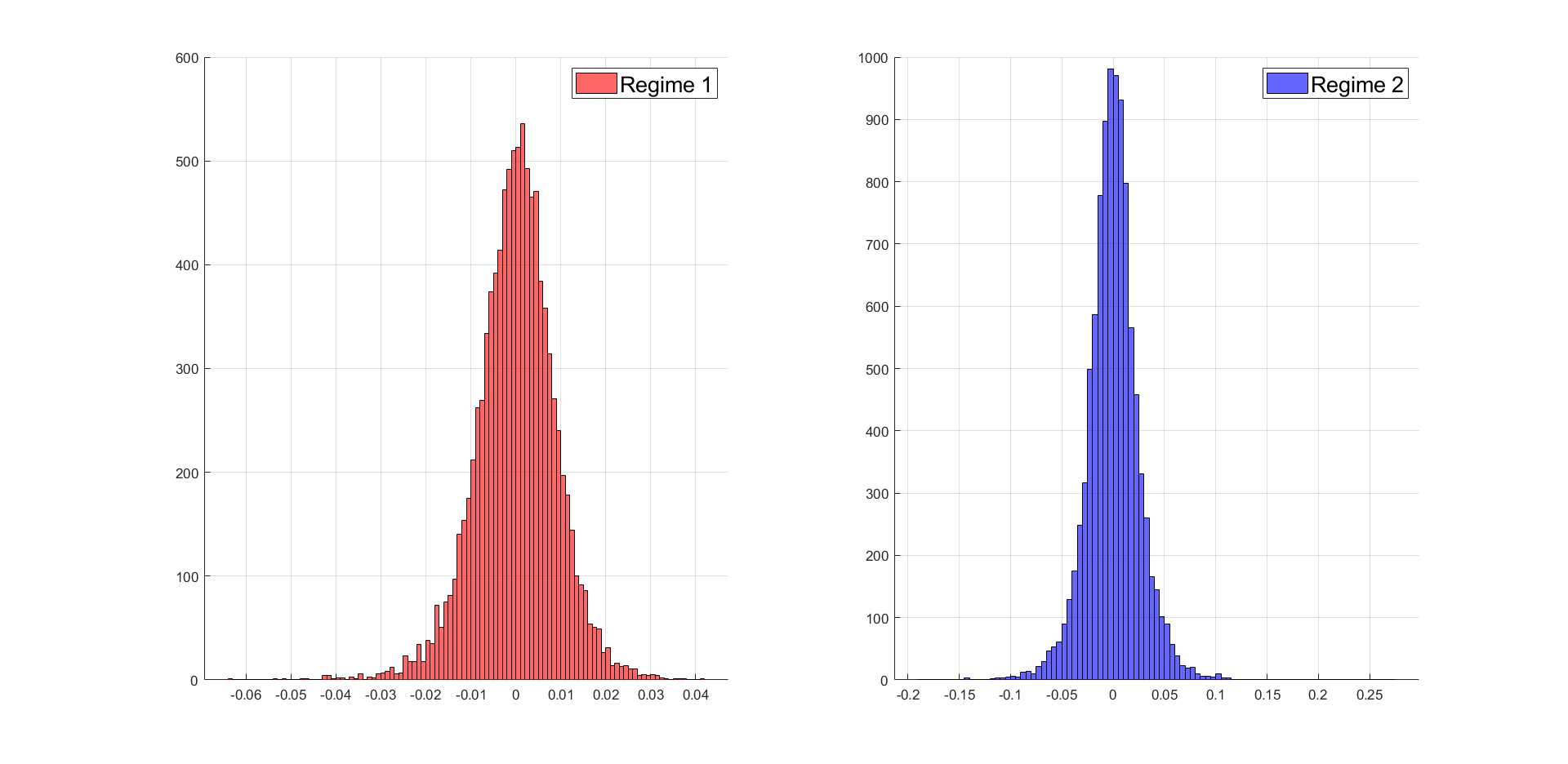}
\caption{Nasdaq regimes, separated.} \label{fig:nas.3}
    \end{minipage}%
    }
    \hfill
    \fbox{%
    \begin{minipage}[t]{0.47\linewidth}
       
\includegraphics[width=1\textwidth]{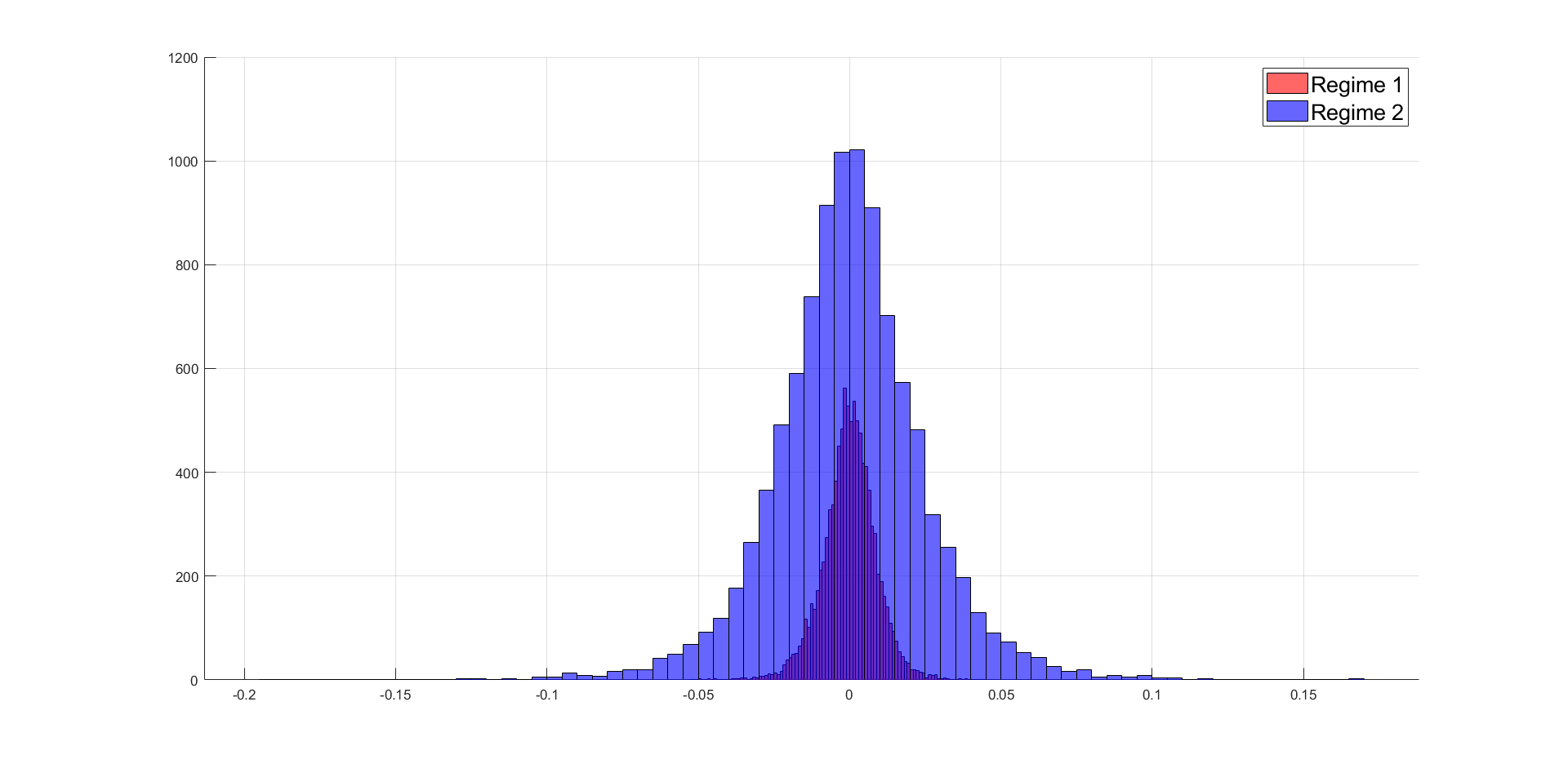}
\caption{Regimes, superimposed.} \label{fig:nas.4}
    \end{minipage}
    }
\end{figure}

We then follow the two-stage estimation strategy proposed by \cite{CG16}. Figures \ref{fig:nas.3} and \ref{fig:nas.4} point to the presence of identifiable regimes.

\section{Results}
\begin{figure}[htbp!]
\centering
\caption{NASDAQ close prices, log returns, and regimes.} 
\label{fig:nas.1}
\centering
\includegraphics[width=1\textwidth]{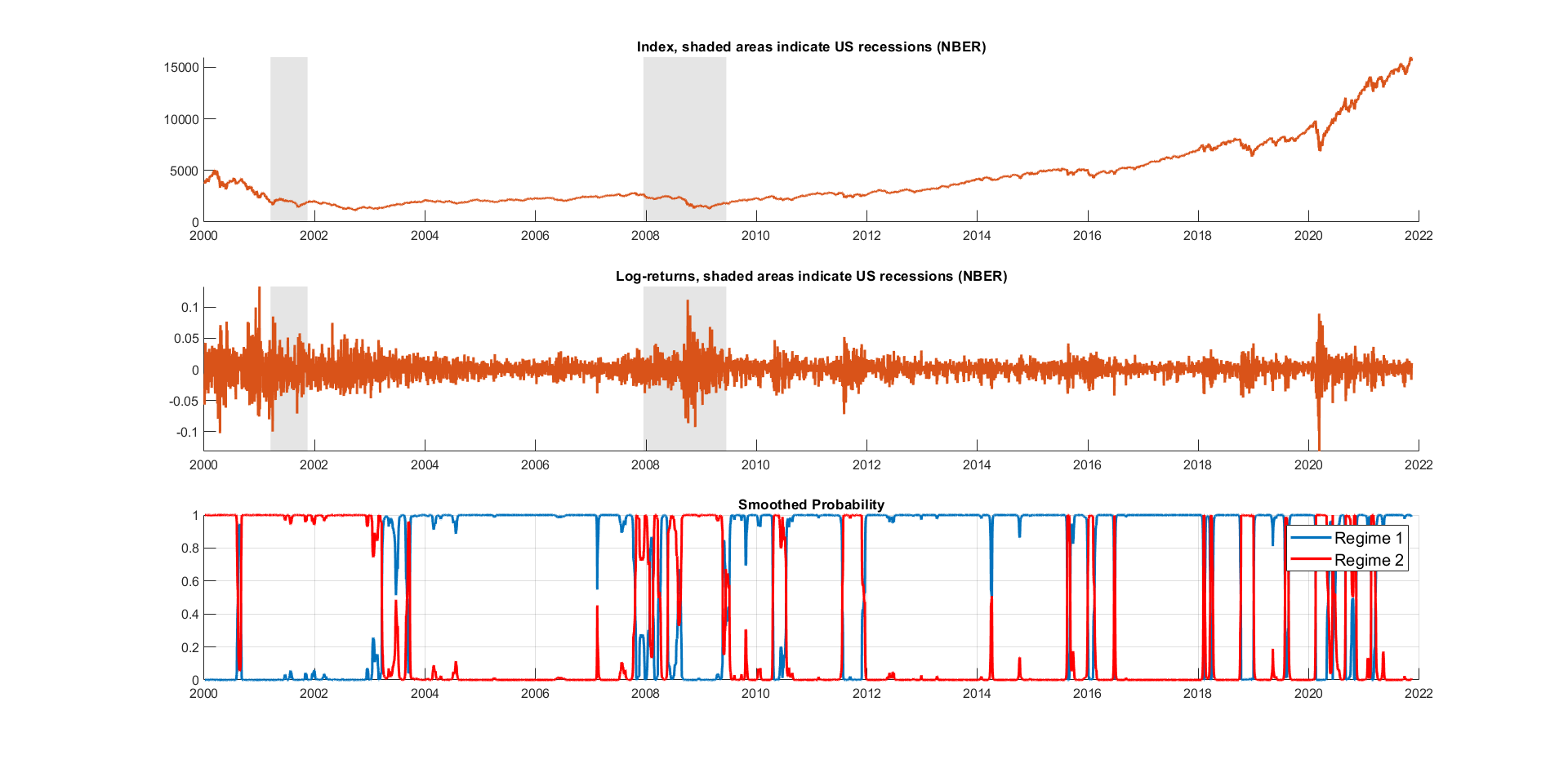}
\end{figure}

Figure \ref{fig:nas.1} shows closing prices for the NASDAQ (top), together with the volatility of log-returns (middle), and the evolution of regimes in our MRS model. Tables \ref{tbl:nig.1} and \ref{tbl:nig.2} report the NIG distribution parameters of the L\'evy jump process fitted to each regime.

\begin{table}[htbp!]
\centering
\caption{NIG distribution parameters of the L\'evy jump process fitted to each regime.} 
\begin{tabular}{lllll}
NIG     & $\alpha$    & $\beta$     & $\delta$    & $\mu$       \\ \hline
State 1 & 150.0919 & -16.2944 & 0.011949 & 0.001276 \\
State 2 & 41.8416  & 0.295358 & 0.026838 & -0.00015
\label{tbl:nig.1}

\end{tabular}
\end{table}

In Table \ref{tbl:nig.1}, the parameter $\alpha$ indictes the jump intensity. The higher the parameter, the lower the jump intensity in a given regime. We observe that $\alpha$ $\approx$ 150 during regime 1, which indicates a low jump intensity. Conversely, $\alpha$ $\approx$ 42 during regime 1 points out a high jump intensity, respectively. $\delta$ indicates the scale parameter representing a measure of the spread of the returns. We can observe that $\delta$ for regime 2 is 0.027 versus 0.12 for regime 1, indicating that the returns are around 2.5 times as dispersed in the second regime compared to the first.

\begin{table}[htbp!]
\centering
\caption{NIG diffusion parameters.} 
\begin{tabular}{lllll}
NIG     & $\kappa$    & $\theta$    & $\sigma$    & $P^Z_{ii}$ \\ \hline
State 1 & 1.011003 & 0.000994 & 0.000082 & 0.716268                 \\
State 2 & 1.082196 & -0.00154 & 0.000643 & 0.283732   
\label{tbl:nig.2}
\end{tabular}
\end{table}

Table \ref{tbl:nig.2} shows the diffusion parameters of the L\'evy jump process fitted to each regime. The mean reverting parameter $\kappa$ is close to unity in both regimes. The volatility parameter $\sigma$ is more than seven times higher during regime 2 than during regime 1. Thus, the substantially higher jump intensity observed in regime 2 translates into increased volatility. The smoothed probabilities in the lower panel of Figure \ref{fig:nas.1} shows the probability of staying in the current regime, which is high for regime 1. This is shown numerically in the last column of Table \ref{tbl:nig.2}: the probability to stay in regime 1 (in regime 2) is equal to 71.6\% (28.3\%). If the chain exhibits some type of general persistence (e.g., a high likelihood of remaining in a particular regime), this might have significant consequences for computing the Value-at-Risk and dynamic portfolio allocation. 

What these graphs and numerical results primarily demonstrate is that the stochastic process suited to each regime does not exhibit the same jump characteristics across the sample period. Indeed, certain eras in the index have a high proportion of jumps (recorded under regime 2), while others do not. As a result, this set of findings demonstrate the merits of using the regime-switching L\'evy model to simulate the price dynamics. Next, we examine the quality of regime classification and model fit.

\subsubsection{Regime Classification Measure of Ang and Bekaert (2002)}
A great model is one that is able to sharply classify the regimes, whilst smoothed probabilities should be either $\approx$ zero or $\approx$ one. To address this question directly, Regime Classification Measures (RCMs) have been proposed by Ang and Bekaert \cite{AB} as a way to determine if the number of regimes $K$ is appropriate. The RCM statistic spans from 0 (perfect regime classification) to 100 (failure to detect any regime classification), and can be formalised as follows:
\begin{equation}
    RCM(K)=100 \times \Big( 1-\frac{K}{K-1}\frac{1}{T}\sum^N_{k=1}  \sum^N_{Z_{t_k}}  \Big( P(Z_{t_k}=i | \mathcal{F}^X_{t_M};\hat{\Theta}_1^{(n)})-\frac{1}{K}      \Big)^2\Big),
\end{equation}
where $P(Z_{t_k}=i | \mathcal{F}^X_{t_M};\hat{\Theta}_1^{(n)})$ corresponds to the smoothed probability and $\hat{\Theta}_1^{(n)}$ is the vector of estimated parameters. $RCM \in [0, 100]$ and lower values are preferred to higher ones. In this sense, a 'perfect' model will be associated with a RCM of almost 0, a good model will have a RCM of close to $\approx$ 0, while a model that cannot distinguish between regimes at all will have a RCM close to 100. A good model is one that implies that the smoothed probability is less than 0.1 or greater than 0.9. This means that the data at time $t\in[0, T]$ is in one of the regimes at the 10\% error level. The RCM was extended for multiple states by Baele \cite{Baele}. 

\subsubsection{Smoothed probability indicator}
The quality of classification may also be observed when the smoothed probability is less than $p$ or greater than $1 - p$ with $p \in [0, 1]$. Thus the data at time $k \in {1, \dots , N}$ has a probability higher than $(100 - 2p)\%$ in one of the regimes for the $2p\%$ error. This per centage is the smoothed probability indicator with $p\%$ error, denoted in Table \ref{tbl:RCM} by $P^{\%}$.

\begin{table}[htbp!]
\centering
\caption{Regime Classification Measure (RCM) and Smoothed Probability Indicator}
\begin{tabular}{lll}
Nasdaq & RCM  & $p^{10}$   \\ \hline
       & 8.63 & 91.39
\label{tbl:RCM}
\end{tabular}
\end{table}

\subsubsection{Measuring the quality of our model's regime qualification performance}
A natural question to ask is how to measure the quality of our model's regime qualification performance. In Table \ref{tbl:RCM} we observe that the RCM statistic for our model is less than 10, indicating a good MRS model fit. We take this as evidence that our model with two regimes is able to provide an advantageous fit to the data. In other words, given the set of models considered in our analysis, and using a reasonably long time span (n=5483 obs), we can conclude that a L\'evy regime-switching model satisfactorily identifies two regimes for the Nasdaq index returns within our sample.


\subsection{Construction of random portfolios}

Following Alexeev and Tapon \cite{ATa12}, we use a simulation approach to construct random portfolios based on actual daily equity returns. Again, we rely on our sample of observations over the period 04 Jan 2000 - 12 Nov 2021. In order to construct our portfolios and depart from the classical Gaussian limitations of \cite{ATa12}, we proceed by drawing random values from the Markov-modulated Normal-Inverse Gaussian (NIG) distribution. The parameters of the model are specified in Tables \ref{tbl:nig.1} and \ref{tbl:nig.2}.

It is important to note that there is a link between security returns and their uncertainty. However, diversification is made possible because of the low or negative correlations between assets. This complicates analysis. It is simple and natural to look at the five variances and ten correlations or covariances if there are only a few securities involved, such as five stocks. As an example, if the number of stocks is 200, then merely looking at the 200 variance and 19900 correlations or covariances will not be particularly useful. Dimension reduction using principal component analysis (PCA) is a well-known statistical technique. In this context, it gives an alternate strategy that can lessen the complexity of our analysis by looking only at certain of the variances and correlations or covariances that we have in our data, but which are supposed to be uncorrelated.

\begin{figure}[htbp!]
\centering
\caption{Weights of our constructed portfolio (5483 observations, 100 assets) based on parameters recovered from market data based on distributional and functional form assumptions.} 
\label{fig:weights}
\centering
\includegraphics[width=1\textwidth]{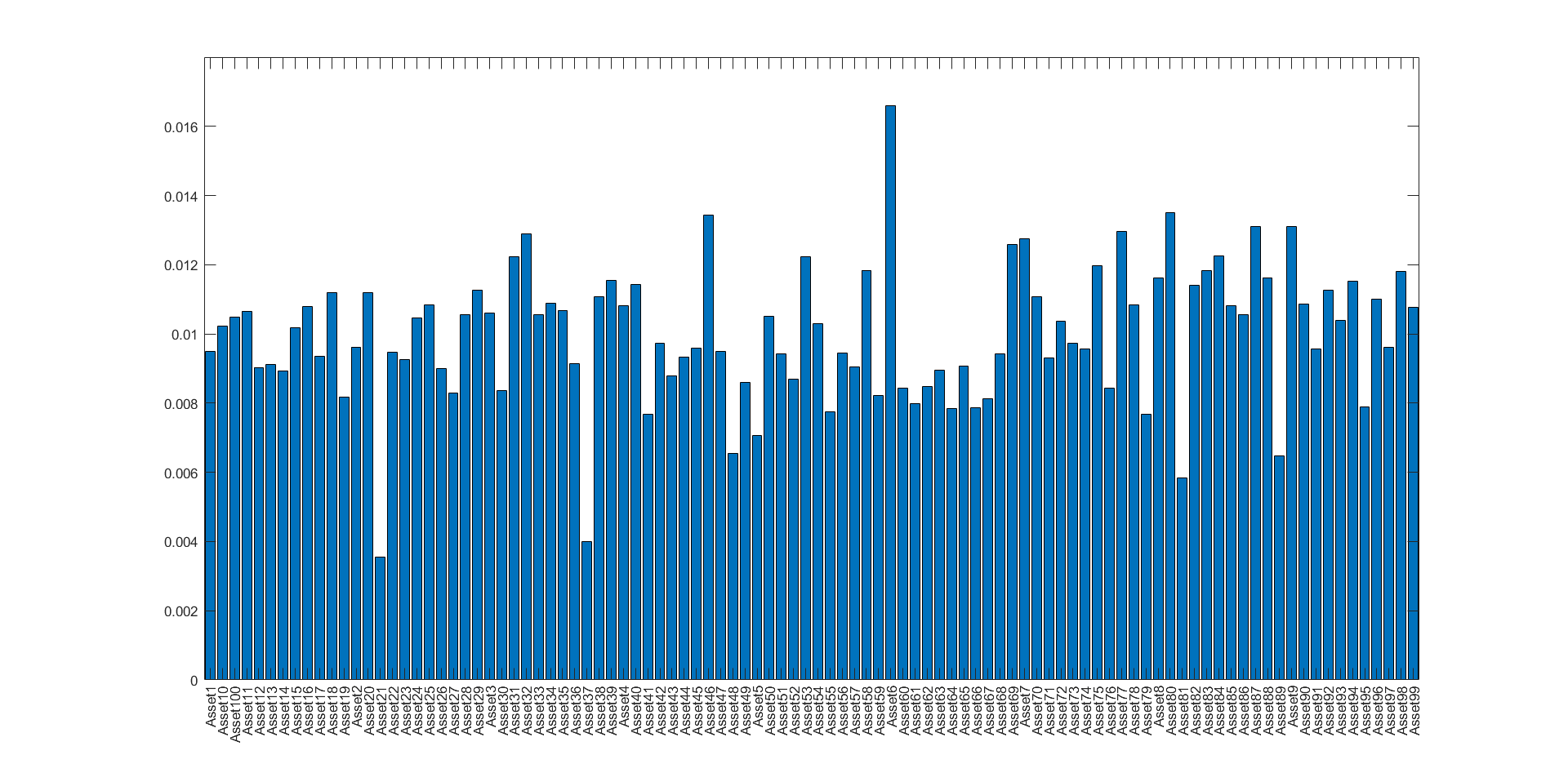}\hfill
\end{figure}

We proceed with our illustrative example based on the methodology suggested in \cite{Meu} (p. 131). A factor model is used to optimise asset allocation in a mean-variance framework. It is common to employ multifactor models for risk modelling, portfolio management, and performance attribution. A multifactor model decreases the size of the investing universe and explains much of the market's unpredictability. Factors might be statistical, macroeconomic, and fundamental. We generate statistical factors from asset returns and then optimise our allocation in accordance with those statistical factors. Figure \ref{fig:weights} shows asset weights obtained from our factor model.

Our results show that 94 (87) components are required to explain at least 95 (90) per cent of variability. These results are in tension with those presented in Alexeev and Tapon \cite{ATa12} and related papers (eg \cite{AD15}), which suggest a significantly smaller sizes of portfolios required for achieving
most diversification benefits. With this information, we may draw three conclusions. In the first place, it may be necessary for the ordinary investor to own a greater number of equities than previously thought. Second, when it comes to portfolio construction, additional work may be needed to improve the estimation of asset returns and integrate some observable aspects in financial time series, in an effort to capture higher moments, regimes, and other such dynamics. And third, the 1/N naive-diversification rule could at least serve as a first clear benchmark for evaluating the success of a certain strategy for optimum asset allocation, presented either by academic research or the investment management sector.

\section{Discussion}
\label{sec:discussion}
\subsection{Dealing with regimes and higher moments}
To determine how many assets are necessary to achieve optimal diversification, there are a number of factors to consider, including: the investment universe (size, asset classes, and features of the asset classes), the investor's characteristics, changes in asset features over time, the model adopted to measure diversification (i.e., equally weighted versus an "optimal" allocation), and the frequency of the data that is being used together to determine how many assets are needed. We argue that it is also important to be mindful of well-evidenced nuances of financial time series data, particular that is possibly non-normally distributed, skewed, multi-regime, leptokurtic distributions or generally non-linear and/or discontinuous price processes.

The cost of overseas investment must also be taken into account when evaluating the advantages of international diversification. When building a portfolio, diversification, security analysis, and asset allocation are some of the challenges that investors face when investing outside national borders. Investment in the home market is not hampered by currency and political concerns, limits on money transfers between nations or various regulations that apply in different countries.

There is a variety of opinions in the academic literature on what constitutes successful diversification. According to some recent research -- see, for example, \cite{ATa12}, \cite{RA21}, \cite{AD15} and the literature cited therein -- the average size of a well-diversified portfolio appears to be greater than it was in previous studies, and this might be attributed to decreased transactional costs in particular. This research also found that the ideal number of equities that compose a well-diversified portfolio increased significantly after taking into account a longer length of time. According to \cite{Z+21} U.S. unsystematic risk appears to have grown dramatically in recent decades relative to the total stock market volatility, underscoring the necessity for bigger portfolios to decrease diversifiable risk to the greatest extent feasible. Many investors may not be able to achieve the same amount of diversity by following the same portfolio size guideline. When making this selection, it is important to consider factors such as how often data is collected; how much risk is taken into account; how confident investors are; and how much diversification benefits may be gained from the selected investment opportunity set, among other things.

\subsection{Simple heuristic diversification rules}
Our results also point to the so called 1/n investment puzzle, whereby some market participants split their contributions evenly among all assets. Since it opposes the core principles of contemporary portfolio theory, it has been suggested that this is a naive method. There is evidence \cite{BT} that suggests that many participants in defined contribution plans allocate their contributions among the different asset classes using basic heuristic diversification methods. The 1/n rule is a common diversification heuristic that is often referred to as a "equally-weighted portfolio". However, it has been criticised for not being on the efficient frontier, as it is not an optimum portfolio. According to some academics, pension systems should be less flexible in order to prevent people from making poor investing decisions.

 However, \cite{Wind} present simple arguments to illustrate that this behaviour may be less naive than it may appear at first and show that the 1/n rule offers certain advantages in terms of robustness. Whilst we do not advocate any heuristic diversification rule in this paper, we note that a certain amount of work may need to go into improving on this simple diversification criterion. In this context, \cite{Job} observe that "naive formation rules such as the equal weight rule can outperform the Markowitz rule." \cite{Mich} also notes that because of estimation risk "an equally weighted portfolio may often be substantially closer to the true MV optimality than an optimized portfolio."

\subsection{Attempts at a unified theory of portfolio management.}
Before concluding our discussion, we ought to briefly discuss the feasibility of applying the MVO framework to assets other than equities, such as fixed income. Any attempt at diversification will involve attempts to diversify across asset classes. However, when it comes to bond portfolios, the established MV approach of portfolio management poses some issues and constructing a theory of bond portfolios is non-trivial. To explain this point in more detail, stocks and bonds differ in numerous ways, the most important of which is the fixed maturity date (i.e. "time of maturity") at which bonds disappear from the market, whereas the characteristics of stocks can fluctuate in response to business news or management decisions. Because the maturity period may take on an infinite number of values in an unconstrained market, there are an infinite number of distinct bonds. Therefore, the price of a stock is only determined by the risks it entails (market risk, idiosyncratic risk). But the price of a bond is determined by both the risks it entails and the amount of time before maturity. The stochastic differential equations used to model stock prices are typically autonomous (meaning that the coefficients are time-independent functions of the prices, as in geometric Brownian motion or mean-reverting processes), whereas any model for bond prices must incorporate the fact that volatility goes to zero when the time to maturity goes to zero, at least this is how it is mathematically expressed. 

A simple portfolio of stocks and bonds is made more analytically difficult by the fact that prices for each kind of asset change according to various principles, even in the most basic scenario. Furthermore, because of the maturity dependency, many techniques that are permissible in the stock market are no longer applicable in the bond market: for example, a simple buy-and-hold strategy results in the conversion of bonds into cash at maturity. And in the context of fixed income derivatives, it is natural to model interest rates with uncountably many traded instruments. Whilst some interesting attempts toward a unified theory of portfolio management, including both stocks and bonds, have been made, the literature on this topic remains relatively sparse and the problem of multi-asset portfolio optimisation remains very much an open research question. In this vein, Ekeland and Taffin \cite{ET05} made attempts to introduce a bond portfolio management theory based on foundations similar to those of stock portfolio management by modelling the discounted price curve as infinite-dimensional dynamics in a Banach space of continuous functions driven by a cylindrical Wiener process. The interested reader is also referred to \cite{Ringer}, who extend these results using a HJM framework, and \cite{Benth}, who apply these results to commodity futures markets. Extending these results could be a fruitful direction for future work.

\section{Conclusion}
\label{sec:conclusion}
The discussion of this paper some of the issues that arise in the portfolio allocation literature. However, we wish to focus the reader's attention on challenges of a different kind, which seem to be absent from the discussions in the optimum portfolio diversification literature. Specifically, there is merit in exploring certain stylized features of financial time series in an attempt to capture dynamics of non-normally distributed, skewed, multi-regime, and leptokurtic asset return distributions.

We promoted the estimation of a Markov-switching model augmented by jumps, under the form of a L\'evy process. After setting up the general structure of L\'evy processes, we outlined their properties with reference to path variation and the L\'evy-Khintchine theorem. The analysis highlighted the properties of Markov chain, such as irreducibility, aperiodicity, and ergodicity. 

By deploying a Markov-modulated L\'evy process, we relaxed a number of assumptions in Alexeev and Tapon \cite{ATa12} in order to account for account for non-normally distributed, skewed, multi-regime, and leptokurtic asset return distributions. We calibrated a regime-switching L\'evy model to equity market data to demonstrate that such a model is a) analytically tractable and computationally effective, b) intuitive, and c) does a good job of matching the empirical moments, including those of higher order. Finally, we argue that a part of the related literature on portfolio diversification relies on assumptions that are in tension with certain observable regularities and which, if ignored, may lead to material underestimation of risk.

\newpage

Code availability: Matlab code is available on request.

\newpage

\section{Technical Appendix}
The following arguments follow directly from \cite{CG16}.
\subsection{Stage 1: The regime-switching model}
We aim to estimate the set of parameters $\Theta=\hat{\Theta}_1:=(\hat{k}_i, \hat{\theta}_i, \hat{\sigma}_i, \hat{\Pi}_i)$ for $i \in S$.
\begin{enumerate}
\item Start with initial vector $\hat{\Theta}_1^{(0)}:=
\Big(\hat{k}_i^{(0)}, \hat{\theta}_i^{(0)}, \hat{\sigma}_i^{(0)}, 
\hat{\Pi}_i^{(0)} \Big)$ for $i \in S$. Let $N \in \mathbb{N}$ be the maximum number of iterations. Fix a positive constant $\epsilon$ as a convergence constant for the estimated log-likelihood function.
\item Assume that we are at the $n+1 \leq N$ steps. Then calculation in the previous iteration of the algorithm gives the following vector set 
$\hat{\Theta}_1^{(n)}:=
\Big(\hat{k}_i^{(n)}, \hat{\theta}_i^{(n)}, \hat{\sigma}_i^{(n)}, 
\hat{\Pi}_i^{(n)} \Big)$
 \end{enumerate}

\subsection{EM-algorithm}
\subsubsection{Expectation step (E step)}
We aim to estimate both filtered probability and smoothed probability. Optimality is achieved when a model is able to identify regimes sharply, such that smoothed probabilities approach either zero or one. Filtered probability is given by the probability such that the Markov chain $Z$ is in regime $i \in S$ at time $t$ with respect to  $\mathcal{F}_T^X$:

For all $i \in S$ and $k=\{1, 2, \ldots, M\}$, estimate the following
\begin{equation}
\begin{split}
& P\Big(Z_{t_k}=i | \mathcal{F}^X_{t_k}; \hat{\Theta}_1^{(n)}\Big)=
\frac{P\Big(Z_{t_k},X_{t_k}|\mathcal{F}_{t_{k-1}}^X;\hat{\Theta}_1^{(n)}\Big)}{f\big(X_{t_k} | \mathcal{F}_{t_{k-1}}^X;\hat{\Theta}_1^{(n)}\Big) } \\
& = \frac{P \Big(Z_{t_k}=i | \mathcal{F}^X_{t_{k-1}}; \hat{\Theta}_1^{(n)}\Big) f \Big(X_{t_k} | Z_{t_k}=i;\mathcal{F}_{t_{k-1}};\hat{\Theta}_1^{(n)} \Big)}{\sum_{j \in S}P \Big(Z_{t_k}=j | \mathcal{F}_{t_{k-1}}^X;\hat{\Theta}_1^{(n)}\Big)  f \Big(X_{t_k} | Z_{t_k}=j;\mathcal{F}_{t_{k-1}};\hat{\Theta}_1^{(n)} \Big)}
\end{split}
\end{equation}
such that 
\begin{equation}
\begin{split}
& P \Big(Z_{t_k}=i | \mathcal{F}^X_{t_{k-1}}; \hat{\Theta}_1^{(n)}\Big) = \sum_{j \in S} P \Big(Z_{t_k}=i, Z_{t_{k-1}}=j | \mathcal{F}^X_{t_{k-1}}; \hat{\Theta}_1^{(n)}\Big) \\
& = \sum_{j \in S} P \Big(Z_{t_k}=i, Z_{t_{k-1}}=j | \hat{\Theta}_1^{(n)}\Big) P \Big( Z_{t_{k-1}}=j | \mathcal{F}^X_{t_{k-1}}; \hat{\Theta}_1^{(n)} \Big) \\
& = \sum_{j \in S} \Pi^{(n)}_{ij} P \Big( Z_{t_{k-1}}=j | \mathcal{F}^X_{t_{k-1}}; \hat{\Theta}_1^{(n)} \Big),
\end{split}
\end{equation}
where $f \Big(X_{t_k} | Z_{t_k}=i;\mathcal{F}_{t_{k-1}};\hat{\Theta}_1^{(n)} \Big)$
 is the density of the process $X$ at time $t_k$, conditional that the process is in regime $i \in S$. Using previous arguments we can observe that, given $\mathcal{F}^X_{t_{k-1}}$, the process $X_{t_k}$ has a conditional Gaussian distribution $\sim N\Big(k_i^{(n)}\theta_i^{(n)}+(1-k_i^{(n)})X_{t_{k-1}},\sigma_i^{2(n)}\big).$ The density of this distribution is given by 
 \begin{equation}
f \Big(X_{t_k} | Z_{t_k}=i;\mathcal{F}_{t_{k-1}};\hat{\Theta}_1^{(n)} \Big)= 
\frac{1}{\sqrt[]{2\pi \sigma_i^{(n)}}} \exp \Big[\frac{X_{t_k}-\big(1-k_i^{(n)})X_{t_{k-1}}-\theta_i^{(n)}k_i^{(n)}\big)^2}{2\big(\sigma_i^{(n)}\big)^2}  \Big]
 \end{equation}
 
On the other hand, to estimate smoothed probability we need to examine when Markov chain $Z$ is in regime $i \in S$ at time $t$ with respect to all the historical data $\mathcal{F}_T^X$. For all $i \in S$ and $k=\{M-1, M-2, \ldots, 1\}$ we obtain 
 \begin{equation}
P\Big(Z_{t_k}=i | \mathcal{F}^X_{t_M}; \hat{\Theta}_1^{(n)}\Big) = 
\sum_{j \in S} \Big(
\frac{P \big(Z_{t_k}=i | \mathcal{F}^X_{t_k}; \hat{\Theta}_1^{(n)}\big) P \big(Z_{t_{k+1}}=j | \mathcal{F}_{t_M};\hat{\Theta}_1^{(n)}|\Pi_{ij}^{(n)} \big)}{P \big(Z_{t_{k+1}}=j | \mathcal{F}_{t_k}^X;\hat{\Theta}_1^{(n)}\big) } \Big)
 \end{equation}
 
\subsubsection{Maximization step (M step)}
We are able to obtain explicit formula of the maximum likelihood estimator of the initial subset of parameters $\hat{\Theta}_1$. The maximum likelihood estimates $\hat{\Theta}^{(n+1)}_1$ for all parameters, for all $i \in S$, can be obtained by
\begin{equation}
\begin{split}
& \theta_i^{(n+1)}= \frac{\sum^M_{k=2} \big[P\big(Z_{t_{k}}=i |\mathcal{F}_{t_M}; \hat{\Theta}_1^{(n)}\big)\big( X_{t_{k}}-(1-k_i^{(n+1)})X_{t_{k-1}}\big]}{k_i^{n+1}\sum^M_{k=2}\big[P\big(Z_{t_{k}}=i |\mathcal{F}_{t_M}; \hat{\Theta}_1^{(n)}\big) \big]} \\
& k_i^{(n+1)}= \frac{\sum^M_{k=2} \big[P\big(Z_{t_{k}}=i |\mathcal{F}_{t_M}; \hat{\Theta}_1^{(n)}\big)X_{t_{k-1}}B_1 \big] }
{
 \sum^M_{k=2} \big[ P \big( Z_{t_k}=i |\mathcal{F}_{t_M}; \hat{\Theta}_1^{(n)}
\big) X_{t_{k-1}}B_2
\big]}  \\
& \sigma_i^{(n+1)}= \frac{
\sum^M_{k=2} \big[ P \big( Z_{t_{k}}=i |\mathcal{F}_{t_M}; \hat{\Theta}_1^{(n)} \big) \big( X_{t_{k}}-k_i^{(n+1)} \theta_i^{(n+1)}  (1-k_i^{(n+1)} ) X_{t_{k-1}}\big)^2 \big] }
{
 \sum^M_{k=2} \big[ P \big( Z_{t_k}=i |\mathcal{F}_{t_M}; \hat{\Theta}_1^{(n)}
\big) \big]}
\end{split}
\end{equation}
where
\begin{equation*}
\begin{split}
& B_1= X_{t_{k}}-X_{t_{k-1}} =
\frac{
\sum^M_{k=2} \big[ P \big( Z_{t_{k}}=i |\mathcal{F}_{t_M}; \hat{\Theta}_1^{(n)} \big) \big(X_{t_{k}}-X_{t_{k-1}} \big) }
{
 \sum^M_{k=2} \big[ P \big( Z_{t_k}=i |\mathcal{F}_{t_M}; \hat{\Theta}_1^{(n)}
\big) \big]}      \\
& B_2= 
\frac{
\sum^M_{k=2} \big[ P \big( Z_{t_{k}}=i |\mathcal{F}_{t_M}; \hat{\Theta}_1^{(n)} \big) X_{t_{k-1}} \big] }
{
 \sum^M_{k=2} \big[ P \big( Z_{t_k}=i |\mathcal{F}_{t_M}; \hat{\Theta}_1^{(n)}
\big) \big]} X_{t_{k-1}}.
 \end{split}
\end{equation*}
 
We then obtain the transition probabilities:
\begin{equation}
\Pi_{ij}^{(n+1)} =
\frac{
\sum^M_{k=2} \Big[ P \big( Z_{t_{k}}=j |\mathcal{F}_{t_M}; \hat{\Theta}_1^{(n)} \big) \frac{\Pi_{ij}^{n}P \big( Z_{t_{k-1}}=i |\mathcal{F}_{t_{k-1}}; \hat{\Theta}_1^{(n)} \big) }{P \big( Z_{t_{k}}=j |\mathcal{F}_{t_{k-1}}; \hat{\Theta}_1^{(n)} \big)} \Big]}
{
 \sum^M_{k=2} \big[ P \big( Z_{t_{k-1}}=i |\mathcal{F}_{t_{k-1}}; \hat{\Theta}_1^{(n)}
\big) \big]}
\end{equation}

Let $\hat{\Theta}_1^{(n+1)}:=({k}_i^{(n+1)}, {\theta}_i^{(n+1)}, {\sigma}_i^{(n+1)}, {\Pi}_i^{(n+1)})$ be the new parameters of the algorithm. These are iterated in step  2 until convergence of the EM algorithm is achieved. The procedure can be stopped if either: 
\begin{itemize}
\item[a)] the procedure has been performed $N$ times; or
\item[b)] the difference between the log-likelihood at step $n+1 \leq N$ and the log-likelihood at step $n$, satisfies the equation 
$logL(n+1)-logL(n)< \epsilon$.
\end{itemize}

Proof of consistency of the (quasi) maximum likelihood estimators is provided in \cite{Kim94}; see also \cite{Ruiz}.
 
\subsection{Stage 2: L\'evy distribution fitted to each regime}
 We have estimated the regime-switching model \ref{SDE} using the EM algorithm. Now, we estimate the set of parameters $\hat{\Theta_2}$ by fitting a NIG distribution for each regime.
\begin{align}
& X(\text{Regime~1}) - L_1(\alpha^1, \beta^1, \delta^1, \mu^1) \\
& X(\text{Regime~2}) - L_2(\alpha^2, \beta^2, \delta^2, \mu^2)
\end{align} 
where $L_1$ and $L_2$ relate to relates to a separate set of Normal Inverse Gaussian distribution parameters of the L\'evy jump process. Estimation of the distribution parameters is done by maximum likelihood, where $\Phi^1=(\alpha^1, \beta^1, \delta^1, \mu^1)$ and $\Phi^1=(\alpha^2, \beta^2, \delta^2, \mu^2)$. Directly following from \cite{CG16}, initialization of the algorithm is performed by the method of moments.

\end{document}